\documentstyle[preprint,aps]{revtex}
\begin{document}
\draft

\title{The Isaacson expansion in quantum cosmology}

\author{Carsten Gundlach}
\address{Department of Physics, University of Utah, Salt Lake City, UT
84112-1195}

\date{10 March, 1993}

\maketitle

\begin{abstract}
This paper is an application of the ideas of the Born-Oppenheimer (or
slow/fast) approximation in molecular physics and of the Isaacson (or
short-wave) approximation in classical gravity to the canonical
quantization of a perturbed minisuperspace model of the kind examined
by Halliwell and Hawking. Its aim is the clarification of the role of
the semiclassical approximation and the
backreaction in such a model.
Approximate solutions of the quantum model are constructed which are not
semiclassical, and semiclassical solutions in which the quantum perturbations
are highly excited.
\end{abstract}


\section{Introduction}

\subsection{Semiclasssical gravity and perturbed minisuperspace}

There is a well-explored and consistent theory of quantized (free)
fields propagating on a fixed classical spacetime \cite{BiDa}.
In particular,
one may construct the expectation value of their stress-energy
tensor. Although it is formally infinite, there is a variety of
regularization procedures which give the same and often sensible
result. A simple example of such a result has
been experimentally verified in the form of the Casimir effect. A more
indirect link of quantum field theory in curved spacetime
to physical reality is
the possibility of  explaining early universe density fluctuations as
quantum fluctuations.

It is tempting to insert the quantum stress-energy tensor
into the classical
Einstein equations which govern the background spacetime. In the
Heisenberg picture we could formally write
\begin{equation}
\label{semi}
G_{ab}[\,g\,]=<\psi|\,T_{ab}[\,g,\hat q\,]\,|\psi>,
\end{equation}
where $g$ is the classical metric, $\hat q$ the quantum field
operators and $|\psi>$ their quantum state.
To make the
theory a consistent approximation, one must include among the quantum
fields the linearized perturbations of the metric itself. Higher
order effects from the quantum nature of the gravitational field
should in contrast be suppressed by a factor of the background curvature scale
over the Planck scale.
It has been
argued that this scheme, called semiclassical gravity, is a better
approximation to the presumed quantum theory of gravity than classical
general relativity if the curvature scale of the background
spacetime is much larger than the Planck length. The question if it
really is must be
considered open in the absence of both experiments and a
quantum theory of gravity. The effective equations of motion for the
background
metric, taking into account the backreaction of the quantum fields,
contain higher than second time derivatives, which give
rise to qualitatively different new solutions. It has been argued that
these are artefacts of the approximation to full quantum gravity and
should be excluded in a consistent way \cite{Sim}.

A tentative model for quantum gravity in which the derivation of
semiclassical gravity as an approximation to a more general theory can
be attempted is the canonical quantization of a minisuperspace model
with generic but linearized perturbations. One such model has been
given by Halliwell and Hawking \cite{HalHaw}, where the minisuperspace
model is a closed Friedmann universe driven by a self-interacting real
scalar field which is homogeneous on the surfaces of homogeneity.

The Halliwell-Hawking model does not seem to be totally consistent:
It has been pointed out \cite{Ryan} that
Halliwell and Hawking miss  some components of the Einstein
equation to a self-consistent order by restricting their ansatz for
the background metric before variation. A related problem is the presence of
linearization instabilities due to the symmetries of the spatial
hypersurfaces \cite{Moncrief,Fay}.
In this paper we shall start from the quantized Halliwell-Hawking
model as it stands in spite of these problems. Our reason for this neglect
is simply that  their
resolution will require the imposition of a finite number of additional
constraints on the wave function, in addition to the infinity that is
already there. Although essential for consistency, these should not
change the physical content of the theory, in which we are interested
here.

Our aim is to reconsider the role of the semiclassical backreaction by
a method which applies to the Halliwell-Hawking model,
but which, in its physical meaning, should generalize to other
perturbed minisuperspace models.

Halliwell
\cite{Hal87,Hal89} and other authors
\cite{Cas,Pad,Paz} have given
a derivation of the equations of semiclassical gravity
which is based on making a single-factor WKB ansatz
for the wave function as in reference \cite{HalHaw}. As we shall see
directly, the backreaction term found this way is small by virtue of
the WKB approximation itself, and the theory obtained
therefore less general than
semiclassical gravity. Semiclassical gravity by naive insertion of the
quantum stress tensor into the Einstein equations would in contrast also
allow for highly excited states of the quantum fields which contribute
a large part or all of the total source term of the Einstein
equations. It is this possibility which we examine here, based on the
Wheeler-deWitt equation of reference \cite{HalHaw}. For simplicity we
consider only pure gravity, so that the Friedmann universe will be
driven by gravitational waves only. Classically, this possibility has
been examined a long time ago \cite{Brill}. The approximation method we
shall use is related both to the Born-Oppenheimer
approximation in molecular  physics \cite{BO} and to the short-wave
approximation
of Isaacson for gravitational waves \cite{Is}. In the remainder of
this introduction we review the previous approach to the backreaction
derivation and then the relevant features of the Born-Oppenheimer and
Isaacson approximations.

\subsection{The Halliwell-Hawking approach to the backreaction}

Let us consider a model in
canonical gravity
where the  Hamiltonian constraint is of the form
\begin{equation}
\label{H}
H(Q,P,q,p)=H_0(Q,P)+H_2(Q,P,q,p)=0,
\end{equation}
where $H_{2}$ is homogeneous of order 2 in the $p$ and $q$.
In reviewing the general way in which Halliwell and Hawking find an
approximate solution to the corresponding Wheeler-deWitt equation, we can limit
ourselves to a toy model which will display the features relevant for
our argument. For simplicity of notation alone we assume therefore that
the quantum operator version of $H_0$ is simply
\begin{equation}
\label{H1}
H_0(Q,-i{\partial\over\partial Q})=-{1\over2}{\partial^2\over \partial
Q^2}+V(Q).
\end{equation}
(We have assumed a positive definite kinetic energy term. In gravity it is
indefinite.)
With the ansatz $\Psi(Q,q)=e^{iS(Q)}\psi(Q,q)$
we then obtain
\begin{equation}
{1\over2}S'^2\psi-iS'\psi'-{1\over2}\psi''-{i\over2}S''\psi+V\psi+H_2\psi=0,
\label{junk}
\end{equation}
where a prime denotes $\partial/\partial Q$ or $d/dQ$. Dividing by
$\psi$ we can separate this equation into
\begin{equation}
{1\over2}S'^2-{i\over2}S''+V+E(Q)=0,
\end{equation}
\begin{equation}
\left[-iS'{\partial\over\partial Q}-{1\over2}{\partial^2\over\partial
Q^2} +H_2-E(Q)\right]\psi=0.
\end{equation}
Here $E(Q)$ is an arbitrary separation function. It has no direct
physical significance, as its choice does not affect the product
$e^{iS}\psi$, but an appropriate choice can help in the interpretation
of the two separate equations of motion.
We now assume that both $e^{iS}$ and the total wave function
$e^{iS}\psi$ are of the WKB form in the variable(s) $Q$, that is we assume
\begin{equation}
|S'^2|\gg|S''|,
\end{equation}
\begin{equation}
|S'|\gg|(\ln \psi)'|.
\end{equation}
To leading order in this approximation
we obtain the Hamilton-Jacobi equation for $S(Q)$,
\begin{equation}
\label{HamJac}
{1\over2}S'^2+V+E(Q)=0.
\end{equation}
By the argument $S'=P=\partial L/\partial \dot Q=\dot Q$
(a dot denotes $\partial/\partial t$ or $d/dt$), we can
identify $S'\partial/\partial Q$ with $\partial/\partial t$ in the WKB
case. To leading order we then obtain a time-dependent Schroedinger
equation \cite{Banks,HalHaw}
\begin{equation}
\label{Sch}
\left[-i{\partial\over\partial t}+H_2-E(Q)\right]\psi=0,
\text{ where }{\partial\over\partial t}\equiv S'{\partial\over\partial Q}.
\end{equation}
One might be tempted to set $E(Q)=<\psi|H_2\psi>$ in (\ref{HamJac})
in order to obtain
the Hamilton-Jacobi form of the semiclassical backreaction (\ref{semi}) in the
Einstein equations in this way, as was once proposed by Hartle
\cite{Har86}. But in (\ref{Sch}) this leads to
$<\psi|-i\partial/\partial t\,\psi>=0$, so that the other part of the
semiclasscial gravity scheme, the time-dependent Schroedinger
equation describing the evolution of the quantum fields, is lost.
If we want to obtain the correct time-dependent
Schroedinger equation for the gravitons, i. e. one where the
Hamiltonian is the usual Hamiltonian of linearized gravitational waves
in the given background $Q(t)$, we must set $E(Q)=0$. But then the
background evolves without a backreaction from the perturbations.
In order to obtain semiclassical gravity as it is commonly understood,
one would have to set $Q(Q)$ equal to $0$ in (\ref{Sch}), and equal to
$<\psi|H_2\psi>$ in (\ref{HamJac}), a question of having the cake and
eating it.

A subtle argument for the presence of a backreaction in spite of
this dilemma has been given by Halliwell \cite{Hal87}. He determines what the
background spacetime really is, not from equation
(\ref{HamJac}),
but
by a Wigner function analysis of the total wave function. (The
convention for $E(Q)$ is then irrelevant.) Following Halliwell,
one may assume that
the perturbations $q$ are for some reason unobservable in principle.
Predictions should then be made, not from the wave function
$\Psi(Q,q)$, but from the density matrix which is obtained by tracing
over the perturbations, $\rho(Q,Q')=\int \Psi(Q,q)^*\,\Psi(Q',q) \,dq$.
The Wigner function derived from this density matrix
may then be positive definite and may be peaked
around a classical trajectory in phase space $(Q,P)$. Halliwell showed
that this trajectory is not given by $P(Q)=S'(Q)$, but rather by
\begin{equation}
\label{P}
P(Q)=S'(Q)+<\psi|-i{\partial\over\partial Q}\psi>.
\end{equation}
Putting this into the Hamilton-Jacobi equation (\ref{HamJac}), using
(\ref{Sch}) and
treating the second term in (\ref{P}) as small, one obtains
\begin{equation}
{1\over2}P^2+V+<\psi|H_2\psi>=0.
\end{equation}
(As predicted, the choice of $E(Q)$ does not matter.) As a way of
finding the correct physics in spite of the appearances of the
equations of motion (\ref{HamJac},\ref{Sch}), this is an impressive
argument.
It is
limited in two ways. By the assumption $|S'|\gg|(\ln \psi)'|$ which
went into its derivation, the backreaction term derived in this way is
necessarily small
and cannot qualitatively change the background. This limitation is
also apparent in the fact that the equations one has to solve are still
(\ref{HamJac}) and (\ref{Sch}), and not those of semiclassical
gravity.

Secondly, only those quantum perturbations contribute to the
backreaction which are by assumption unobservable. There are good
tentative reasons for this assumption.  In a closed universe which has
undergone inflation, for example, one would not be able to observe the
quantum state of the perturbations completely because of the presence
of particle horizons \cite{Hal89}. But why should one only ever be
able to observe the gravitational backreaction of those quantum fields
that are ``out of sight''?

The limitation of the backreaction effect to a small perturbation of the
background trajectory  is even less satisfactory.
One should expect to find a quantum
equivalent of the classical situation where the self-gravity of
gravitational waves substantially curves the background spacetime on
which they move. Examples of such spacetimes which have been examined
are a nearly static spherically symmetric concentration of gravitational waves
(a gravitational geon \cite{BriHar}), or a Friedmann universe closed by
and driven by gravitational waves \cite{Brill}.

\subsection{The classical Isaacson approximation}

A key ingredient for both these cases was supplied by Isaacson
\cite{Is}
(see also \cite{Bur}). He considered a one-parameter family of solutions of the
vacuum Einstein equations of the form
\begin{equation}
\label{g}
g_{ab}(x;\epsilon)=\gamma_{ab}(x)+h_{ab}(x;\epsilon)
\end{equation}
and assumed that the family of perturbations $h_{ab}(x;\epsilon)$ is such that
their amplitude scales like $\epsilon$, but their ``wavelength'' like
$\epsilon^{-1}$. Formally, one assumes that there is a coordinate
system in which
\begin{equation}
\label{significance}
h_{ab}(x;\epsilon)=O(\epsilon),\quad h_{ab,c}(x;\epsilon)=O(1),\quad
h_{ab,cd}(x;\epsilon)=O(\epsilon^{-1}),
\end{equation}
where a comma denotes a partial (coordinate) derivative. $\gamma_{ab}$ and its
derivatives are considered as $O(1)$.
Let the Ricci tensor of $g_{ab}$  be expanded in powers of $h_{ab}$:
\begin{equation}
R_{ab}(g_{ab},g_{ab,c},g_{ab,cd})=R_{ab}^{0}(\gamma)+R_{ab}^{1}(\gamma,h)
+R_{ab}^{2}(\gamma,h)
+\ldots{}=0,
\end{equation}
where $R_{ab}^{1}$ is linear in $h_{ab}$, $R_{ab}^{2}$ quadratic, and so on.
Here $R_{ab}^{1}$ is a contraction (with $\gamma^{ab}$)
of $\nabla_{a}\nabla_{b}h_{cd}$, where $\nabla_{a}\gamma_{bc}\equiv0$, and
$R_{ab}^{2}$ is a contraction of $\nabla_{a}\nabla_{b}h_{cd}h_{ef}$ .
Inserting the ansatz (\ref{g}) into the vacuum Einstein equations, and
separating powers of $\epsilon$, one
obtains to leading and next order
\begin{equation}
\label{lin}
R_{ab}^{1}(\gamma)=0
\end{equation}
\begin{equation}
\label{quad}
R_{ab}^{0}(\gamma)=-<R_{ab}^{2}(\gamma,h)>_{\rm average}
\end{equation}
(Strictly speaking, the connection terms in $R_{ab}^{1}$
and $R_{ab}^{2}$ are of a
lower order than the partial derivative terms and do not appear
in these equations). The brackets in (\ref{quad}) (and here only)
refer to a suitable spacetime averaging.
The right-hand side of (\ref{quad}) is an effective stress tensor of
the perturbations which drives the background. Equation (\ref{lin}) is
a linear wave equation for the perturbations propagating on the
background. These equations have to be solved together in a
self-consistent way, and as such are not part of a perturbation scheme
in which each order  is solved before and independently
of the higher ones. The difference from conventional perturbation
theory is also seen from the fact that the limit $\epsilon\to 0$ is singular.

\subsection{The Born-Oppenheimer approximation}

It is worth reviewing the Born-Opppenheimer approximation in a
notation which is sufficiently abstract to be suggestive both of its
original context of molecule physics and of the context it will be
used in in this paper. Let us assume that we
are dealing with a Hamiltonian of the form
\begin{equation}
\label{noP}
H(Q,P,q,p)=H_0(Q,P)+H_2(Q,q,p).
\end{equation}
The single but crucial difference to (\ref{H}) is that $H_2$ depends only on
$Q$, but not on the conjugate momentum (or momenta) $P$.
We
want to solve the
time-independent Schroedinger equation
\begin{equation}
\label{tiSch}
H(Q,-i{\partial\over\partial Q},q,-i{\partial\over\partial
q})\Psi(Q,q)=E\Psi(Q,q),
\end{equation}
where $E$ is a given constant. (The case of canonical quantum gravity
differs
from this only
by the presence of additional constraints
on the wave function, and by the restriction $E=0$.)
We use the fact that $H_2$ does not contain any $Q$-derivatives to
pose the eigenvalue problem
\begin{equation}
\label{eigenvalue}
H_2(Q,q,-i{\partial\over\partial q})f_\nu(Q,q)=E_\nu(Q)f_\nu(Q,q)
\end{equation}
for the complete sets $f_\nu$ and $E_\nu$, for each value of the
parameter $Q$. We choose the $f_\nu$ to be orthonormalized
under the scalar product
\begin{equation}
\label{scalprod}
<f_\nu|f_\mu>=\int f_\nu^* f_\mu dq.
\end{equation}
(They are of course already orthogonal for $E_\nu\neq E_\mu$.)
We now make the ansatz
\begin{equation}
\label{prod}
\Psi(Q,q)=\sum_\nu\Psi_\nu(Q)f_\nu(Q,q)
\end{equation}
and solve for the $\Psi_\nu$.
Let us again assume, for simplicity of notation, that $H_0$
is of the form (\ref{H1}).
(It would be straightforward to write down what follows for $H_0$ being
an arbitrary second order derivative operator in more than one
variable. This general case includes canonical gravity.)
We can simplify the result obtained from substituting (\ref{prod})
into (\ref{tiSch}) by
splitting it into components with respect to the basis $f_\nu$ using
the scalar product (\ref{scalprod}).
We obtain an infinite number (labelled by $\nu$) of coupled equations
in the minisuperspace variable(s) $Q$:
\begin{equation}
\label{molecule}
\left(H_0+E_\nu(Q)-E\right)\Psi_\nu(Q)=
\sum_\mu\left({1\over2}<f_\nu|{\partial^2\over\partial Q^2}f_\mu>\Psi_\mu
+<f_\nu|{\partial\over\partial Q}f_\mu>{\partial\over\partial
Q}\Psi_\mu\right)
\end{equation}
In the molecular physics application, where $Q$ are the nucleus
positions and $q$ the electron positions, it can be shown that the
right-hand side is small by a factor of the square root of the ratio
of the electron mass over the proton mass. The argument itself does not seem
capable of extension to gravity, although it has been argued that
the Planck mass must be involved in finding a small number in the
problem.
This is obviously
inapplicable to the case of pure gravity, where the Planck mass is the
only fundamental scale. In our
particular example of a perturbed Friedmann universe
the suppression factor will turn out to be the
ratio of the length scales of the metric perturbations over the
Hubble scale of the
background metric, a very intuitive result given the
usual language in which the $q$ are called the fast variables and the $Q$ the
slow variables.
Neglecting the right-hand side coupling terms in (\ref{molecule}),
and assuming $E=0$, we
can
suggestively
write it as
\begin{equation}
\label{suggestive}
\left(H_0+<f_\nu|H_2f_\nu>\right)\Psi_\nu(Q)=0
\end{equation}
Here we seem to have obtained the backreaction equation
we were looking for in a direct manner and without making any
arbitrary choices. However, we are not really solving two equations
self-consistently, in the manner of (\ref{lin}) and (\ref{quad}).
Instead we have to do more, namely solve (\ref{eigenvalue}) for all
values of $Q$. There is no classical equivalent of this.

In section II we describe the Hamiltonian formulation of a perturbed
minisuperspace model -- that of Halliwell and Hawking without matter --
which is is completely described by a Hamiltonian constraint of the
form (\ref{noP}). We make the first steps towards the Born-Oppenheimer
approximation by solving the fast-part eigenvalue problem
(\ref{eigenvalue}) and then writing down the slow-part equation
(\ref{molecule}).  In section III we find the Born-Oppenheimer
approximation to that model, by finding the regime in which the
right-hand side of (\ref{molecule}) can be neglected. This turns out
to be the Isaacson approximation.  In section IV we recover
semiclassical gravity by adding the WKB assumption. We compare our
results with previous work. Section V resumes our results and raises
some questions on the validity of our starting point
(equation (\ref{Wada}) below) and the posssibility of generalizing our
treatment to other perturbed minisuperspace models.

\section{Born-Oppenheimer treatment of a perturbed minisuperspace model}

Perhaps the simplest physically complete perturbative minisuperspace
model is that of a Friedmann universe without matter, but with generic
(small) perturbations of the metric. For a closed ($K=1$) Friedmann
universe this is a special case of the model of Halliwell and Hawking
\cite{HalHaw}. There is a single Hamiltonian constraint of the form
(\ref{H}) -- where $H_{2}$ is the integral over all space of the
quadratic Hamiltonian constraint, and $H_{0}$ is the minisuperspace
Hamiltonian constraint -- and an infinity of constraints (the
linearized Hamiltonian and momentum constraints at each space point)
which are homogenous of order 1 in the $p$ and $q$. (There should also
be a finite number of additional constraints which are homogeneous of order 2,
but as stated in the introduction, with Halliwell and Hawking we
neglect these here.)

A canonical transformation that both solves the linearized
Hamiltonian and momentum constraints and brings the remaining part of the
Hamiltonian constraint into the ``Born-Oppenheimer" form
(\ref{noP}) has been given by Wada
\cite{Wada86}.
The wavefunction then depends only on some of the new variables, and
is subject only to a single Hamiltonian constraint. The remaining
degrees of freedom in this
particular model are
the amplitudes
of  transverse traceless perturbations of the 3-metric, $d_n$, where $n$ labels
the space dependence (the $n$th tensor harmonic on the 3-sphere),
and the minisuperspace variable $\tilde
\alpha$, which is equal to $\alpha$, the logarithm of the scale factor, up to
terms quadratic in the perturbations.

Although physically
interesting in its own right,
this model will here also serve as a toy model for
a more general
perturbative minisuperspace ansatz. Therefore we use the previous
notation $Q$ for the minisuperspace variables  and
$\vec q=\{q_{3},q_{4},\ldots{}q_{n},\ldots{}\}$ for the perturbation
variables. ($n$ is an infinite discrete label which arises from
splitting the perturbations into ``Fourier'' components with respect
to their spatial dependence.)
In the following
$Q$ is therefore the same as $\tilde\alpha$ and $q_n$ is the same as
$d_n$ of
reference
\cite{Wada86}, where their precise definitions can be found. The
definition of the perturbed spacetime metric in terms of these variables is
given in \cite{HalHaw}.
For understanding the physics of this particular model it suffices to
know that
the background metric is (now using our notation)
\begin{equation}
\label{background}
ds^2=-N_0^2 dt^2 + e^{2Q} d\Omega^2_3,
\end{equation}
where $d\Omega^2_3$ is the round metric on the unit three-sphere, and
where $N_0$ and $Q$ are arbitrary functions of $t$,
and that $q_{n}$ is the amplitude of a transverse traceless
perturbation of the 3-metric of comoving wavelength $\sim 1/n$. The total
Hamiltonian of the perturbed system, in the remaining degrees of freedom
given by
Wada (but in our notation)
is
\begin{equation}
\label{Wada}
H=N_0e^{-3Q}\left[-{1\over2}P^2-e^{4Q}+{1\over2}\sum_{n=3}^\infty\left(p_n^2+
(n^2-1)\,e^{4Q}\,q_n^2\right)\right].
\end{equation}
It is constrained to vanish.
We shall find  a family of approximate solutions -- in the Isaacson
limit -- of the corresponding Wheeler-deWitt equation
\begin{equation}
\label{WdW}
\left[{1\over2}{\partial^2\over\partial Q^2}-e^{4Q}
+{1\over2}\sum_{n=3}^\infty\left(-{\partial^2\over\partial q^2}+
(n^2-1)\,e^{4Q}\,q_n^2\right)\right]\Psi(Q,\vec q)=0.
\end{equation}
We have made an ad-hoc choice of factor ordering.

As a first step, as yet without an approximation, we split the
Wheeler-de-Witt equation in an infinity of variables into an infinity
of equations in only the minisuperspace variables, here only $Q$. This
is precisely the ansatz that prepares the Born-Oppenheimer
approximation in molecule physics. We make the ansatz
\begin{equation}
\label{prodfield}
\Psi(Q,\vec q)=\sum_{\vec \nu}\Phi_{\vec \nu}(Q)\prod_{n=3}^\infty
f_{\nu_n}\left((n^2-1)^{1/4}e^Qq_n\right).
\end{equation}
Here $\vec\nu=\{\nu_3,\nu_4,\nu_5,...,\nu_n,...\}$ is an infinite tupel
of numbers which have range $\nu=0,1,2,...$.
$f_\nu(x)$ is, for
each value of $\nu$, a given function of its argument $x$, namely the
normalized energy eigenfunction of the unit frequency harmonic
oscillator, defined by
$(1/2)(-d^2/dx^2+x^2)f_\nu(x)=(\nu+1/2)f_\nu(x)$
and $\int f_\nu^2\,dx=1$.
In this definition we have already incorporated the solution of the eigenvalue
equation (\ref{eigenvalue}) for $H_{2}$ given by (\ref{WdW}).
As
the $f_\nu$
form a basis of square-integrable functions on the real line, our
ansatz (\ref{prodfield}) is generic.
We note that $d/dQ f_\nu=x\, d/dx f_\nu$ where $x$ is the
formal argument of $f_{\nu_n}$, namely
$(n^2-1)^{1/4}e^Qq_n$.
We introduce the operators $a=(x+d/dx)/\sqrt{2}$,
$a^{\dag}=(x-d/dx)/\sqrt{2}$ and $N=a^{\dag}a$,
with the well-known properties $af_\nu=\sqrt{\nu}\,f_{\nu-1}$,
$a^{\dag}f_\nu=\sqrt{\nu+1}\, f_{\nu+1}$, $af_0=0$, and $Nf_\nu=\nu f_\nu$.
Now we follow the Born-Oppenheimer
procedure of separating the $\Phi_{\vec\nu}$ by the orthonormality of
the
$f_\nu$. For this purpose we
use the identity
$x\, d/dx= (1/2)(a^2-a^{\dag
2}-1)$ and its square $(x\, d/dx)^{2}=(1/4)(a^4+a^{\dag 4}-2a^2+2a^{\dag
2}-2N^2-2N-1)$.
The final result is
\begin{eqnarray}
\label{longone}
\nonumber\left( {1\over2}{d^2\over dQ^2}-e^4Q
+e^{2Q}\sum_n(\nu_n+{1\over2})\sqrt{n^2-1}
\right)\Phi_{\vec\nu}
=-{1\over2}\sum_{n}\Bigl[
\tilde\Phi_{\vec\nu|\nu_{n}+2}'
-\tilde\Phi_{\vec\nu|\nu_{n}-2}'
-\Phi_{\vec\nu}'
\Bigr]\\
\nonumber-{1\over8}\sum_{n}\Bigl[
\tilde\Phi_{\vec\nu|\nu_{n}+4}
+\tilde\Phi_{\vec\nu|\nu_{n}-4}
-2\tilde\Phi_{\vec\nu|\nu_{n}+2}
+2\tilde\Phi_{\vec\nu|\nu_{n}-2}
-(2\nu_{n}^{2}+\nu_{n}+1)\Phi_{\vec\nu}
\Bigr]\\
\nonumber-{1\over8}\sum_{n}\sum_{m\neq n}\Bigl[
\tilde\Phi_{\vec\nu|\nu_{n}+2,\nu_{m}+2}
+\tilde\Phi_{\vec\nu|\nu_{n}-2,\nu_{m}-2}
-2\tilde\Phi_{\vec\nu|\nu_{n}-2,\nu_{m}+2}
-2\tilde\Phi_{\vec\nu|\nu_{n}+2}
+2\tilde\Phi_{\vec\nu|\nu_{n}-2}
+\Phi_{\vec\nu}
\Bigr].\\
\end{eqnarray}
Again, a prime denotes $\partial/\partial Q$ or $d/dQ$.
To keep this expression readable, we have introduced a shorthand notation
that implies certain factors arising from the Bose-Einstein
statistics of the excitations of the perturbation modes $q_{n}$. It is
most easily explained by example:
\begin{eqnarray}
\label{a}
\tilde\Phi_{\vec\nu|\nu_{n}-4}
&&=\sqrt{n_{\nu}}\sqrt{n_{\nu}-1}\sqrt{n_{\nu}-2}\sqrt{n_{\nu}-3}
\Phi_{\{\nu_3,\nu_4,...,\nu_{n-1},\nu_n-4,\nu_{n+1},...\}},\\
\label{b}
\tilde\Phi_{\vec\nu|\nu_{n}+2}'
&&=\sqrt{n_{\nu}+2}\sqrt{n_{\nu}+1}{d\over dQ}
\Phi_{\{\nu_3,\nu_4,...,\nu_{n-1},\nu_n+2,\nu_{n+1},...\}}\text{ etc.}\\
\nonumber
\end{eqnarray}

The left-hand side of equation (\ref{longone}) contains
an infinity of the form $\sum_{n=3}^{\infty}\sqrt{n^{2}-1}$,
and the right-hand side several infinities of the form
$\sum_{n=3}^{\infty}1$.
We must eliminate these if the equation is to have more than
formal meaning. A slightly more consistent procedure than to just
cross out these terms is to normal-order all occurences of $a$ and
$a^{\dag}$,
i. e. to put
$a$ to the right of $a^{\dag}$ in all products. In $H_{2}$
we obtain $N$ instead of $N+1/2$. This is in
analogy to
standard regularization procedure for free quantum field theory in flat
space. $x\,d/dx$ becomes $:x\,d/dx:=(1/2)(a^{2}-a^{\dag 2})$.
Replacing $(x\,d/dx)^{2}$ by $:(x\,d/dx)^{2}:$ does not get rid of all
infinities. Instead we choose
\begin{equation}
\label{choice}
:(:x\, d/dx:)^{2}:=(1/4)(a^4+a^{\dag 4}-2N(N-1)).
\end{equation}
In support of this choice one might argue that $x\,d/dx$ should always be
normal-ordered as a block  as in (\ref{choice}), because it arises from the
expression $\partial f_{\nu}/\partial Q$. With that choice of
normal ordering we obtain
\begin{eqnarray}
\label{normal}
\nonumber\left( {1\over2}{d^2\over dQ^2}-e^4Q
+e^{2Q}\sum_n\nu_n\sqrt{n^2-1}
\right)\Phi_{\vec\nu}
=-{1\over2}\sum_{n}\Bigl[
\tilde\Phi_{\vec\nu|\nu_{n}+2}'
-\tilde\Phi_{\vec\nu|\nu_{n}-2}'
\Bigr]\\
\nonumber-{1\over8}\sum_{n}\Bigl[
\tilde\Phi_{\vec\nu|\nu_{n}+4}
+\tilde\Phi_{\vec\nu|\nu_{n}-4}
-2\nu_{n}(\nu_{n}-1)\Phi_{\vec\nu}
\Bigr]\\
\nonumber-{1\over8}\sum_{n}\sum_{m\neq n}\Bigl[
\tilde\Phi_{\vec\nu|\nu_{n}+2,\nu_{m}+2}
+\tilde\Phi_{\vec\nu|\nu_{n}-2,\nu_{m}-2}
-2\tilde\Phi_{\vec\nu|\nu_{n}-2,\nu_{m}+2}
\Bigr].\\
\end{eqnarray}
In the following we use this equation as our starting point.
We shall refer to the
$\nu_n$ as ``occupation numbers'' by analogy with the states of a set
of harmonic oscillators.
Furthermore we shall
by the same analogy
refer to the first line of the right-hand side of (\ref{normal}) as the
``two-particle terms'' and to the remainder as the
``four-particle terms''.

\section{Quantum version of the Isaacson expansion {\it is} the
Born-Oppenheimer approximation}

We now introduce the Isaacson expansion
as a means of making the right-hand side of (\ref{normal}) small
compared to the last term on the left-hand side,
thus completing a Born-Oppenheimer approximation scheme.
(To post-Born-Oppenheimer order we shall recover semiclassical gravity.)
With Isaacson we
consider a family of approximate solutions labeled by a small
parameter $\epsilon$, such that the large-scale metric, in our case
the Friedmann background, is independent of $\epsilon$. Both the
amplitude and the wavelength of the perturbations are chosen to scale
like $\epsilon$, so that their effective energy-momentum tensor,
suitably averaged, is independent of $\epsilon$ and can act as the
source for the background curvature. In our model,
defined by the Hamiltonian (\ref{Wada}), we identify as a guidance principle
the perturbation wavelength with $1/n$ and the classical amplitude with the
occupation number $\nu_n$. The total energy of the  perturbations is
then proportional to
\begin{equation}
\sum_{n=3}^\infty\nu_n \sqrt{n^{2}-1}\sim\int_0^\infty
\nu(n)\,n\,dn.
\end{equation}
This is of course  the term appearing on the left-hand side
of (\ref{normal}) which we would like to retain as a backreaction term
after the model of (\ref{suggestive}).
To make this expression (approximately) independent of $\epsilon$, we
may consider the
one-parameter family of occupation numbers (classically, amplitudes)
\begin{equation}
\nu_{n}^{(\epsilon)}=\epsilon^2\,\mu(\epsilon n;\epsilon).
\end{equation}
The explicit $\epsilon$-dependence in the second argument of the
function $\mu(n;\epsilon)$ is necessary to make the perturbation energy
precisely $\epsilon$-independent with
this ansatz. Isaacson however uses the fact that for small $\epsilon$
this dependence becomes weak, so that for sufficiently small
$\epsilon$ the function $\mu(n,0)$ characterizes a whole family of
approximate solutions $\nu_{n}^{(\epsilon)}$ by the scaling relation
$\nu_{n}^{(\epsilon)}\simeq\epsilon^2\,\mu(\epsilon n,0)$.

The equivalent in the quantum theory as defined by (\ref{normal}) is a
one-parameter family of wave functions
$\Phi_{\vec\nu}^{(\epsilon)}(Q)$. We define it as
\begin{equation}
\label{PhiF}
\Phi^{(\epsilon)}_{\vec\nu}(Q)=F_{\vec\mu}(Q;\epsilon)
\text{ where } \mu_n=\epsilon^{-2}\nu_{\epsilon^{-1}n}.
\end{equation}
To keep both the value and the suffix of $\mu_{n}$ integer one must
restrict the range of $\epsilon$ to
$\epsilon=N^{-1}$, with $N$ any positive integer.
A similar problem, that the family of approximate solutions cannot be
continuous in $\epsilon$, would arise also in the classical Isaacson
approach to a closed universe. It therefore does not invalidate the existence
of the approximation. In the case of open spatial
hypersurfaces, the suffix $n$
becomes continuous anyway, and one may then simply consider
$F(\vec\mu)$ as a smooth functional on a continuous field.
Alternatively one may set $\epsilon=1$ at
the end of a formal expansion in $\epsilon$. Then one no longer considers a
one-parameter family of solution, but rather a single solution whose
occupation numbers are nonvanishing for  high
frequencies only, and then small.

By substituting the ansatz (\ref{PhiF}) into the
equations of motion for $\Phi_{\vec\nu}$, and then changing variable from
$\vec\nu$ to $\vec\mu$ and summation index from $n$ to $\epsilon n$,
we obtain an equation of motion for $F$ that contains explicit powers
of $\epsilon$ in its coefficients:
\begin{eqnarray}
\label{Fepsilon}
\nonumber\left( {1\over2}{d^2\over dQ^2}-e^4Q
+e^{2Q}\sum_n\nu_n\sqrt{n^2-1}
\right)F_{\vec\mu}(Q;\epsilon)\\
\nonumber
=-{1\over2}\epsilon^{2}\sum_{n}\Bigl[
\sqrt{\mu_{\epsilon n}+2\epsilon^{-2}}
\sqrt{\mu_{\epsilon n}+\epsilon^{-2}}
F'_{\vec\mu|\mu_{\epsilon n}+2\epsilon^{-2}}(Q;\epsilon)
-\sqrt{\mu_{\epsilon n}}
\sqrt{\mu_{\epsilon n}-\epsilon^{-2}}
F'_{\vec\mu|\mu_{\epsilon n}-2\epsilon^{-2}}(Q;\epsilon)
\Bigr]\\
\nonumber-{1\over8}\epsilon^{4}\sum_{n}\Bigl[
\sqrt{\mu_{\epsilon n}+4\epsilon^{-2}}
\sqrt{\mu_{\epsilon n}+3\epsilon^{-2}}
\sqrt{\mu_{\epsilon n}+2\epsilon^{-2}}
\sqrt{\mu_{\epsilon n}+\epsilon^{-2}}
F_{\vec\mu|\mu_{\epsilon n}+4\epsilon^{-2}}(Q;\epsilon)
\Bigr]
-{1\over8}\epsilon^{4}\sum_{n}\sum_{m\neq n}\Bigl[\ldots{}
\Bigr].\\
\end{eqnarray}
We have not written out all terms. The remainder can easily be
reconstructed from (\ref{normal}). The important point is that
all ``two-particle'' terms of the right-hand side are
multiplied by a factor of $\epsilon^{2}$ and all ``four-particle terms''
by a factor of $\epsilon^{4}$. These factors arise from
the Bose statistics prefactors, which we have written out here for this reason.

We see from the presence of these explicit factors of $\epsilon$ in
(\ref{Fepsilon}) that it was necessary to have
given $F$ an explicit dependence on $\epsilon$ (as well as the
dependence of its formal argument $\vec\mu$ on $\epsilon$).
But by analogy with the work of
Isaacson, we expect that solutions for different values of
$\epsilon$ are related by a simple scaling of their arguments, as $\epsilon$
becomes small. Expecting the
same physical phenomenon in the quantum theory, we have already
incorporated that scaling behaviour into (\ref{PhiF}). We now express our
expectation that the explicit $\epsilon$-dependence of $F$ disappears as
$\epsilon \to 0$ by expanding it as
\begin{equation}
\label{expansion}
F_{\vec\mu}(Q;\epsilon)
=\chi_{\vec\mu}(Q)\,\rho_{\vec\mu}(\epsilon^{2} Q)\sigma_{\vec\mu}
(\epsilon^4 Q)...
\end{equation}
The final form of the equations of motion with our ansatz is then,
after separating powers of $\epsilon$ and then setting $\epsilon=1$,
\begin{equation}
\label{chi}
\left({1\over2}{d^2\over dQ^2}
+e^{2Q}\sum_n \nu_n\sqrt{n^2-1}-e^{4Q}
\right)\chi_{\vec\nu}(Q)=0,
\end{equation}
\begin{equation}\label{rho}
\chi_{\vec\nu}'(Q)\,{d\over dQ}\rho_{\vec\nu}(Q)
=-{1\over2}\sum_n\left(
\chi_{\vec\nu|\nu_n+2}'(Q)\,
\tilde\rho_{\vec\nu|\nu_n+2}(Q)
-\chi_{\vec\nu|\nu_n-2}'(Q)\,
\tilde\rho_{\vec\nu|\nu_n-2}(Q)\right),\quad{\rm etc.}
\end{equation}

The leading order equation (\ref{chi}) is of course just (\ref{Fepsilon}) with
$\epsilon$
set
equal to zero. As such it is the
Born-Oppenheimer approximation to the quantum equations of motion.
{}From the analogy with the classical
Isaacson expansion we expect that it  describes
the motion of the background driven by the averaged perturbations.
The next order
(\ref{rho}) should describe ``the motion of the
perturbations
on that
background''. In particular, if the solution to (\ref{chi}) is of the
WKB form, thus implying a classical
background $Q(t)$, we should expect (\ref{rho})
to describe a free quantum field theory on the
curved classical spacetime given by (\ref{background}).
The following order, a first-order linear differential equation for
$\sigma$
with coefficients depending on $\chi$ and $\rho$,
should describe higher order corrections that we
do not expect to understand in this intuitive sense, and we
have therefore not written it out.
It may be worth clarifying that, although the physical significance of
$\epsilon$ here is the same as in (\ref{significance}), mathematically
it leads to a perturbation expansion as in (\ref{HamJac},\ref{Sch})
rather than a self-consistent
field expansion as in (\ref{lin},\ref{quad}).

Before we examine the
semiclasscical gravity interpretation of (\ref{chi}), (\ref{rho}),
we note here that this set of equations has solutions which are
definitely not semiclassical. We have therefore found approximate
solutions of the
pure gravity Halliwell-Hawking model which are not semiclassical, as
promised in the abstract. In other words, it does not actually make a
difference to solving the perturbation equations if the background is
behaving classically or not, because we need not look for a
time-dependent Schroedinger equation.

\section{WKB approximation and
semiclassical gravity}

We now compare our results with the corresponding quantum field theory
of perturbations on a fixed background spacetime, with a given $Q(t)$
(for a given choice of $N_{0}(t))$ in the background metric
(\ref{background}). As a first step we must formulate precisely what
we mean by quantum field theory in a curved spacetime in the context
of our model of a Friedmann universe with gravitational wave
perturbations. The Friedmann universe (\ref{background}) is to be the
classical background, and the metric perturbations are to be
the quantum fields.
As we want to make contact with quantum cosmology, we choose the
Schroedinger picture.

Instead of a Wheeler-deWitt equation the perturbations obey the
time-dependent
Schroedinger equation
\begin{equation}
\left[-i{\partial\over\partial t}
+{1\over2}N_0(t)e^{-3Q(t)}
\sum_{n=3}^\infty\left(-{\partial^2\over\partial q^2}+
(n^2-1)\,e^{4Q(t)}\,q_n^2\right)\right]\Psi(t,\vec q)=0.
\end{equation}
We again split the wave function into harmonic oscillator components
\begin{equation}
\Psi(t,\vec q)=\sum_{\vec \nu}\Phi_{\vec \nu}(t)\prod_{n=3}^\infty
f_{\nu_n}\left((n^2-1)^{1/4}e^{Q(t)}q_n\right)
\end{equation}
with the same notational conventions as before.
We may then apply the same procedure of identifying the action of
$\partial/\partial Q$ on the $f_{\nu_n}$ with the action of
annihilation and creation operators. The only difference is that
$\partial/\partial Q$ arises here from $-i\partial/\partial t$
acting on the $Q(t)$ inside the $f_{\nu_n}$. Therefore only first
$Q$-derivatives appear, and the equations of
motion for the coefficients are much simpler. We use normal ordering
again as a regularization procedure. As there are no  terms quartic in
$a$ and $a^{\dag}$, it
is unambiguous. We obtain
\begin{equation}\label{timedep}
\left(-i{d\over dt}+N_0(t)e^{-Q(t)}
\sum_n\nu_n\sqrt{n^2-1}
\right)\Phi_{\vec\nu}
={i\over2}{dQ\over dt}\sum_n\left(
\tilde\Phi_{\vec\nu|\nu_n+2}
-\tilde\Phi_{\vec\nu|\nu_n-2}\right)
\\
\end{equation}

In these equations we recognize the well-known fact that ``particles''
are created in pairs
(due to the symmetry of the background spacetime)
and mainly in spacetime regions where
the curvature scale is less than or comparable to their Compton
wavelength. This is not the place to review definitions of the
particle concept in curved spacetime. It will suffice to consider a
simple situation in which $Q(t)$ is constant, then changes, then is
constant again. In the regions of constant $Q$ we have a well-defined
notion of ground state and particles. The ground state is
that in which all $\Phi_{\vec\nu}$ vanish apart from $\Phi_{\vec 0}$.
If we start with this state at an early time, then after the period of
nonvanishing $\dot Q$,
at late times, the other components $\Phi_{\vec\nu}$ will
no longer vanish. The components with $\nu_n=2,4,\ldots{}$ for a given $n$ will
be excited by being  strongly coupled to the $\vec\nu=\vec 0$ component,
if $|dQ/dt|>N_0e^{-Q}n$ during that period. But $e^Q/n$
is the
physical wavelength of the perturbation labelled by $n$,
and $N_0^{-1}\,dQ/dt$ the Hubble constant (with respect to proper
time). Therefore the criterium that particles with comoving wavelength $1/n$
be produced abundantly
is just that their physical wavelength be greater than the Hubble
distance.

It is illuminating to
separate off the trivial
part of the time evolution
of the $\Phi_{\vec\nu}$ (that is the part which is independent of
their initial values):
\begin{equation}
\label{trivial}
\Phi_{\vec\nu}(t)=\rho_{\vec\nu}(t)\,\exp\left(-i
\sum_n\nu_n\sqrt{n^2-1}
\int^{t} N_0(t')e^{-Q(t')} \,dt'\right)
\end{equation}
($\rho_{\vec\nu}$ defined here is not the same as in equation
(\ref{expansion}), but we have chosen this notation consciously, as
the two definitions will be seen to be related.)
The resulting equation for the $\rho_{\vec\nu}$ is now real:
\begin{equation}
\label{QFT}
{d\over dt}\rho_{\vec\nu}(t)
=-{1\over2}{dQ\over dt}\sum_n\left(
\tilde\rho_{\vec\nu|\nu_n+2}
-\tilde\rho_{\vec\nu|\nu_n-2}\right)
\end{equation}
One way of looking at this equation is as an interaction picture of
quantum field theory: The trivial part of the dynamics has been
suppressed and only the interaction part is described explicitly. In
flat space quantum field theory this is the nonlinear part of the
equation of motion, but in curved space also the interaction with the
background curvature, even for linear fields.

Although equation (\ref{QFT}) is real, it still describes the entire
evolution of the quantum fields $q$ on the background spacetime
(\ref{background}) -- the multiplication by the factor (\ref{trivial})
is merely algebraic. But we have now formulated the time-dependent
Schroedinger equation in a way which evokes Hartle's original attempt
at finding the backreaction in an unexpected manner.
In particular, if we define
$\psi(\vec q,t)=\sum_{\vec\nu}\rho_{\vec\nu}f_{\vec\nu}$,
we find $<\psi|\partial/\partial t\psi>=0$. The energy of the
perturbation has been hidden away in equation (\ref{trivial}),
and we can always reconstruct it trivially.

In
order to establish the relation of (\ref{QFT}) to the full
Wheeler-deWitt
equation,
we can use
$Q(t)$ locally as the independent variable instead of $t$, by
replacing $d/dt$ by $(dQ/dt)d/dQ$. The equation for $\rho_{\vec\nu}$ we obtain
is then very similar to equation (\ref{rho}):
\begin{equation}
\label{similar}
{d\over dQ}\rho_{\vec\nu}(Q)
=-{1\over2}\sum_n\left(
\tilde\rho_{\vec\nu|\nu_n+2}
-\tilde\rho_{\vec\nu|\nu_n-2}\right)
\end{equation}

This resemblance is rather surprising, as we have
not yet introduced a semiclassical approximation, only the Isaacson
expansion.
What is missing to recover (\ref{timedep}) from (\ref{rho})?
The first requirement is that $\chi_{\vec\nu}'\simeq\chi_{\vec\nu|\nu_{n}\pm
2}'$ for all $n$, so that these factors cancel in (\ref{rho}),
allowing us to go from there to (\ref{similar}).
Clearly this is possible for a certain interval of $Q$ given appropriate
initial conditions, because the equations (\ref{chi}) for
$\chi_{\vec\nu}$ and $\chi_{\vec\nu|\nu_{n}\pm 2}$ have nearly the same
coefficients: The difference is only the energy of two gravitons compared
to the energy of all gravitons in the
universe.
Our first requirement is therefore that $\chi_{\vec\nu}(Q)$ be a smooth
function not only of $Q$, but also of $\vec\nu$.

The second requirement, for going from (\ref{similar}) to (\ref{QFT}),
is not surprising: Clearly $dQ/dt$ is not defined for arbitrary
$\chi_{\vec\nu}(Q)$, but only by the
Banks\cite{Banks}-Halliwell-Hawking\cite{HalHaw} approximation.
To obtain it, we must assume that the $\chi_{\vec\nu}$
are of WKB form. Finally, we obtain (\ref{timedep}) from (\ref{QFT}) by the
trivial step (\ref{trivial}).

What is the equation of motion for the semiclassical background? We
substitute the WKB ansatz $\chi_{\vec\nu}=e^{-iS(Q)}$ into the
minisuperspace equation (\ref{chi}) and in the WKB approximation
$|S'|^{2}\gg|S''|$ approximate $\chi''$ by $-S'^{2}\chi$ and
interpret $S'$ as the classical momentum $P$. From the Hamiltonian
(\ref{Wada}) we read off $\dot Q=-N_{0}e^{-3Q}P$ and obtain finally
the classical energy constraint
\begin{equation}
{1\over2}\left({\dot Q\over N_{0}}\right)^{2}+e^{-2Q}=
e^{-4Q}\sum_{n}\nu_{n}\sqrt{n^{2}-1}.
\end{equation}
As $e^{Q(t)}$ is the Friedmann scale factor, and $\dot Q/N_{0}$ the
Hubble constant, this is just the Friedmann equation for a closed
universe in the presence of a radiation fluid, whose density scales as
(scale factor)$^{-4}$.

In rederiving semiclassical gravity we have not quite arrived at (\ref{semi}).
{}From a naive interpretation of that equation one might expect that the
backreaction term of a mixed state is just the sum of the backreaction
terms of its pure-state components. This seems to be true in our
model only if the components have approximately the same energy.
In other words we cannot have the background driven by an effective
perturbation stress tensor which is an average of the stress tensors
of two different pure quantum states, unless these stress tensors are
nearly the same.
Presumably such a situation would not make experimental sense.

It should also be noted that we have found neither negative energy in
the effective stress tensor of the quantum fields -- it is just that
of a radiation fluid -- nor fourth-order effective background
equations. This may be due to our choice of model to start from.

\section{Conclusions}

We have found a more general way of (approximately) solving the
Wheeler-deWitt equation of the Halliwell-Hawking model
than the one making a WKB ansatz for the background variables. Our
approximation requires that the system splits into a slow and fast
part, namely the Friedmann background metric and short-wave metric
perturbations, with intermediate wavelengths in or close to their
ground states. Physically, this assumption corresponds to the Isaacson
approximation in classical gravity, mathematically to the
Born-Oppenheimer approximation in quantum mechanics.

Under this condition the Wheeler-deWitt equation splits
into a system of uncoupled
minisuperspace Wheeler-deWitt equations for the
Friedmann background and a system of coupled first order equations for
the perturbations, with coefficients depending on the solutions of the
background equations. The essential difference to the
Halliwell-Hawking solution
scheme is that one does not have to assume that the background is in a
WKB state to obtain this split.

The WKB form of the background can be chosen as
a second, and independent assumption, which allows to interpret
the minisuperspace WdW equations as a Hamilton-Jacobi equations for a
classical background driven by a classical radiation fluid,
and the system of perturbation equations as a
time-dependent Schroedinger equation for QFT on that classical
background. In other words, the WKB approximation works in the same
way in our scheme as in the Halliwell-Hawking scheme, but the class of
semiclassical solutions we obtain is more general: The Hamilton-Jacobi
equation (the Friedmann equation in our model) contains the energy of
a classical radiation field. This term, the semiclassical backreaction
of the quantum metric perturbations, can be large, i. e. the quantum
fields can be highly excited.

The most important remaining question is whether a Born-Oppenheimer
expansion can be applied to any other perturbed minisuperspace models.
We do not know why the canonical transformation
described by Wada \cite{Wada86} manages to solve all the linearized
constraints of
the Halliwell-Hawking model restricted to pure gravity and at the same
time to bring
the Hamiltonian into a form that allows our Born-Oppenheimer
treatment.
The same kind of transformation, applied by Shirai and Wada
\cite {ShiWa}  to the
full model including scalar field matter, is not so successful: The linearized
Hamiltonian constraint is not solved, and the
zero-plus-second-order Hamiltonian is not of
Born-Oppenheimer form in the remaining degrees of freedom.
There is no obvious reason for this failure.

Wada's treatment is  hard to follow in geometrical
terms, as he works from the Hamiltonian of Halliwell and Hawking,
which is complicated for two different reasons.
Firstly, Halliwell and Hawking split
all perturbations into harmonics on $S^{3}$ from the start. Secondly,
they truncate the action and then calculate the Hamiltonian. One would
expect more insight from truncating the full Hamiltonian directly, a
procedure which may be inequivalent if the background is not itself a
solution.
In truncating the full Hamiltonian directly, the source of terms of
the form $f(Q)Ppq$ and
$f(Q)PPqq$ in $H_{2}$ is clear: They arise from expanding the
``supermetric times momentum times momentum'' part of the full Hamiltonian
constraint.
Still, one may able to see for the general case of a homogeneous
background
if a canonical transformation cannot be found which gets rid of these
terms.
At least one should be able to formulate Wada's transformation in
geometrical terms and explain why it does not generalize, if indeed it
does not.
In such a treatment it would be necessary and probably illuminating to
correct
the inconsistencies of the Halliwell-Hawking model we mentioned in
the introduction.

After this work was completed, reference \cite{DoLa} was brought to
the author's attention. There a general, exact solution of the quantum
problem we have considered here is given. If one makes the canonical
transformation $\bar q_n=(\exp Q)q_n$, $\bar Q=Q+{1\over2}\sum_n
q_n^2$, the Hamiltonian (\ref{Wada}) transforms into a sum of simple
harmonic oscillator Hamiltonians, the one representing $\exp \bar Q$
with a negative sign. (The second part of this transformation is not
made explicit in \cite{DoLa}.) One can therefore give the general
exact solution of the corresponding Wheeler-deWitt equation in terms
of products of harmonic oscillator eigenfunctions.

The simple expression for the transformed Hamiltonian arises in the
approximation in which terms of order $\sum_nq_n^2$ are kept, but
terms of order $(\sum_nq_n^2)^2$ are neglected. This is the same
approximation that was used in truncating the full Hamiltonian in the
first place, i. e. to keep terms to quadratic order in the metric
perturbations. It breaks down towards the initial or the final
singularity or both of any classical solution, where the perturbations
become large and must be treated nonlinearly.

The difference between $\exp Q$ and $\exp \bar Q$ is of order
$\sum_nq_n^2$ and therefore of physical significance: Whereas the
original Hamiltonian of the Friedmann universe of scale factor $\exp
Q$ driven by gravitons is the same as that of a Friedmann universe of
the same scale factor driven by a massless minimally coupled scalar
field, the transformed Hamiltonian is the same as that of a Friedmann
universe with scale factor $\exp
\bar Q$ driven by a conformally coupled scalar or an electromagnetic
field.  A minimally coupled and a conformally coupled massless scalar
field -- or equivalently gravitational and electromagnetic radiation
-- driving a Friedmann background do not give rise to the same
spacetime, beyond leading order in the Isaacson approximation.  Only
in the second case the effective stress-tensor is precisely traceless.
To interpret a classical or quantum solution of the transformed system
one must transform it back to variables $\exp Q$ and $q_n$ (when the
closed form of the solution is lost).  The fact that the background
spacetime with scale factor $\exp Q$ is more than just one of several
equivalent choices of variable is obscured by the simplicity of our
model: In a more complete model the physical scale factor could be
determined with the help of any additional test matter field
introducing a fundamental scale.  The present paper derives solutions
of the quantum system directly in the physical variables. This scheme
is directly linked to the derivation of semiclassical gravity, while
the solution of \cite{DoLa} has the merit of being in closed form.

The ``Born-Oppenheimer" solution scheme might have been introduced by
starting with the canonical transformation $\bar q_n=(\exp Q) q_n$,
$\bar Q=Q$, giving rise to terms $\bar P\bar p\bar q$ and $\bar p\bar
q\bar p\bar q$ in the transformed Hamiltonian. These are just the 2-
and 4-particle interactions we have described. Seen this way, the
4-particle interactions should be neglected as being of quartic order
in the perturbations. This means that we cannot consistently derive
the equation of motion for $\sigma_{\vec\nu}$ and higher order
corrections in $\epsilon$. The author takes this as an indication that
the physical content of a quantized perturbed minisuperspace model
does not go beyond semiclassical gravity -- higher order perturbative
approximations to quantum gravity require a better approximation to
the full classical action.

The author would like to thank Fay Dowker and Raymond Laflamme for
helpful communications.

\thanks

The author would like to thank Karel Kucha\u r for his suggestion that
a quantum equivalent of the gravitational geon should exist, as well
as for his invaluable encouragement and questions.  This work was
supported in part by grant NSF PHY92-07225 and by research funds of
the University of Utah.

\end{document}